# Modeling Physical/Digital Systems: Formal Event-B vs. Diagrammatic Thinging Machine

**Sabah Al-Fedaghi**
*sabah.alfedaghi@ku.edu.kw*
Computer Engineering Department, Kuwait University, Kuwait

**Summary**
Models are centrally important in many scientific fields. A model is a representation of a selected part of the world, which is the model's target system. Here, a system consists of a software portion as a component among many others. Event-B is a modeling method for formalizing and developing systems whose components can be modeled based on set theory and first-order logic. The thinging machine (TM) is a diagram-based model establishes three levels of representation: (1) a static structural description, which is constructed upon the flow of things in five generic operations (activities; i.e., create, process, release, transfer, and receive); (2) a dynamic representation, which identifies hierarchies of events based on five generic events; and (3) a behavioral representation according to the chronology of events. This paper is an exercise in contrasting the formal Event-B to the diagrammatic TM. The purpose is to further understand modeling in computer science. This is motivated by the claim that computer scientists should not invent specific languages to do the modeling. Important notions such as events and behavior are contrasted, and a case study system of traffic on a bridge is modeled in Event-B and TM. The results seem to indicate the need for both modeling approaches.

## 1. Introduction

Models are centrally important to many scientific fields, including models that represent a selected part of the world, which is the model's target system [1]. Such representational models seem to comprise different styles (e.g., mathematical models and data models). They can be considered as structures with mappings between the model and the target system [2]. Usually, a model has its own language, which contains symbols that are interpreted as referring to a structure's objects, relations, or functions [1].

According to Abrial [3], physicists who construct models use classical set-theoretic notations and never invent specific languages to model their systems. Abrial [3] continues by claiming that computer scientists believe it is necessary to invent specific languages to do the modeling and this, according to Abrial [3], "is an error" because set theoretic notations are well suited to perform system modeling in computer science. Abrial's [3] intention is to build a general system within which is a certain piece of software, which is a component among many others. The aim is to construct a complete model of the target system, including the software and its physical environment. Abrial [3] affirms, "We propose to do it by constructing *mathematical models* which will be analyzed by doing proofs." The result of this methodology is Event-B.

This paper is an exercise of contrasting this formal Event-B with a non-formal model called the thinging machine (TM), which is based on diagrams, to discover what has motivated computer scientists to adopt a non-formal approaches to modeling.

### 1.1 Event-B

Event-B [2] is a modeling methodology for formalizing and developing general systems (including software) whose components can be modeled based on set theory and first-order logic. Event-B is currently centered on the general notion of events [4]. This event-based scheme is useful in requirements analysis, modeling of distributed systems, and in the discovery/design of distributed and sequential programming algorithms [5].

Event-B is described as formalism that is "relatively simple, but not very expressive… [in] real-life systems: [this requires] a lot of details [and] huge formal specifications, which are hard to maintain" [6]. The diagrammatic language UML-B (based on UML) has been proposed to complement modeling in Event-B. It combines state machine refinement with class refinement techniques [7]. Four kind of diagrams are provided: package, context, class, and state diagrams. A package diagram is a top-level diagram that shows the structure and relationships between components in a project [7]. UML state diagrams can be used to generate an Event-B specification [6].

**Example**: Consider the state diagram shown in Fig. 1 [6]. The Event-B specification includes the axioms

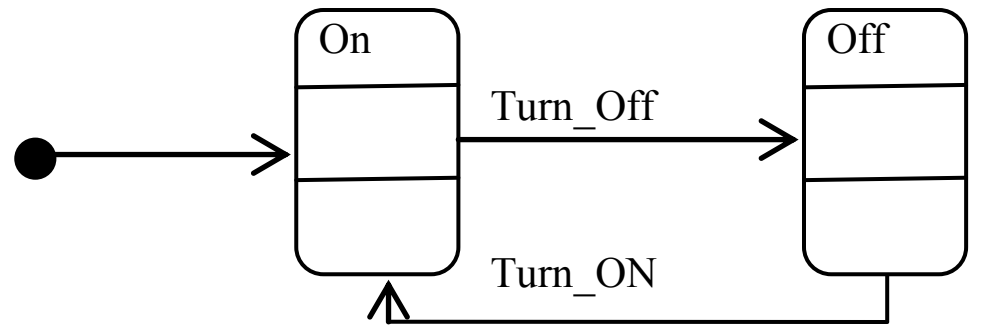

Fig. 1. Sample state diagram

typeof_On: ON ε trafficlight_states
typeof_Off: Off ε trafficlight_states
distinct_states_in_trafficlight_STATES:
partition (trafficlight_STATES, {On}, {off})

The specification of Event-B events for the state diagram includes the following:

```
EVENTS
      INITIALIZATION
           STATUS
       ordinary
BEGIN
   Init_trafficlight :  trafficlight On
END
TurnOn
   STATUS
ordinary
WHEN
    Isin_Off: trafficlight = Off
…
```

### 1.2 Thinging Machine

The TM model is a domain independent conceptual modeling tool that establishes three levels of representation: (1) a static structural description, which is constructed upon the flow of things in five generic operations (activities; i.e., create, process, release, transfer, and receive); (2) a dynamic representation, which identifies hierarchies of events based on the five generic events; and (3) a chronology of events. To achieve a self-contained paper and because TM is a new model that is not widely known, we will review the basic concepts of TM in section 2.

### 1.3 About this Paper

Our aim is to further the understanding of modeling in computer science. Important notions such as events and behavior are contrasted in terms of how each model conceives of these notions, that is Event-B and TM. The results seem to indicate the need for both approaches to modeling.

The next section reviews the basic philosophy of TM. The example in that section is a new contribution. Section 2 discusses the Event-B development approach (e.g., its refinement and proof). This aspect is important because we aim to develop a TM version of an Event-B model. The remaining sections of the paper involve modeling a system of traffic on a bridge, which is specified in Event-B in detail (74 pages) [3]. The system is remodeled in TM.

## 2. Modeling Development

In Event-B, a model is developed through refinement, which is used to relate the abstract model of a system to another, more concrete model while maintaining the same properties of the abstract model [8]. A refinement technique allows the modeler to focus on different aspects of the system at different abstraction levels [7]. According to Abrial [3], a large system has to be modeled in successive steps. Each of these steps makes the model richer by creating and then enriching the various components. In remodeling the bridge example, we will follow this refinement process because the Event-B model of the bridge is developed according to this process.

In Event-B, modeling also consists of proving that the representation fulfills certain desired properties. Proof-based development methods integrate formal proof techniques into system development. Starting with an abstract model, details are added by building a sequence of more concrete ones [5]. This proof-based development is one of the strong aspects of Event-B. In our development of the TM version of the bridge model, we will ignore this feature to limit the focus when contrasting the two models because TM is underdeveloped in this area.

## 3. Thinging Model Theory

This section will briefly review TM modeling to provide a base for this paper's aim of applying TM to the Event-B bridge system. A more elaborate discussion of TM's philosophical foundations can be found in [9-17].

### 3.1 Basics of the Thinging Machine

Typically, ontology requires classifications such as a functional classification of human bodily functions: mental, sensory, speech, respiratory, digestive, and so on [18]. Yet, even with the impressive progress in developing ontologies of things (i.e., entities, objects), the ontology of processes (TM machines) is still a problem [18]. In TM, ontology is based on a single category of entities called thimacs (*thi*ngs/*ma*chines). The thimac is simultaneously an "object" (called a *thing*) and a "process" (called a *machine*)—thus, the name thimac. The thimac notion is not new. In physics, subatomic entities must be regarded as both particles and waves to fully describe and explain observed phenomena [19]. According to Sfard [20], abstract notions can be conceived of in two fundamentally different ways: structurally, as objects/things (static constructs), and operationally, as processes. Thus, distinguishing between form and content and between process and object is popular, but, “like waves and particles, they have to be united in order to appreciate light” [21]. TM adopts this notion of duality in conceptual modeling and generalizes it beyond mathematics.

The term “thing” relies more on Heidegger’s [22] notion of “things” than it does on the classical notion of objects. According to Heidegger [22], a thing is self-sustained, self-supporting, or independent—something that stands on its own. More importantly, it is that which can be spoken about, “that which can be talked about [or] that which is named” [23]. “Talking about” a thing denotes being modeled in terms of being created, processed (change), released, transferred, and/or received. According to Johnson [23], “there is *no* thing that we cannot speak about.” In Heidegger’s [22] words, a thing “things”. That is, it ties together its constituents in the same way that a bridge unifies environmental aspects (e.g., a stream, its banks, and the surrounding landscape). In our TM ontology of dual being, the thing’s machine (the machine side of the thing) “machines”; that is, it operates on (other) things by creating, processing, releasing, transferring, and/or receiving them.

The term “machine” refers to a special abstract machine called a TM (see Fig. 2). The TM is built under the postulation that it only performs five generic operations: creating, processing (changing), releasing, transferring, and receiving.

A thimac has dual being as a thing and as a machine. A thing is created, processed, released, transferred, and/or received. A machine creates, processes, releases, transfers, and/or receives things. We will alternate among the terms “thimac”, “thing”, and “machine” according to the context.

The five TM operations (also called stages) form the foundation for thimacs. Among the five stages, flow (a solid arrow in Fig. 2) signifies conceptual movement from one machine to another or among a machine’s stages. The TM’s stages can be described as follows.

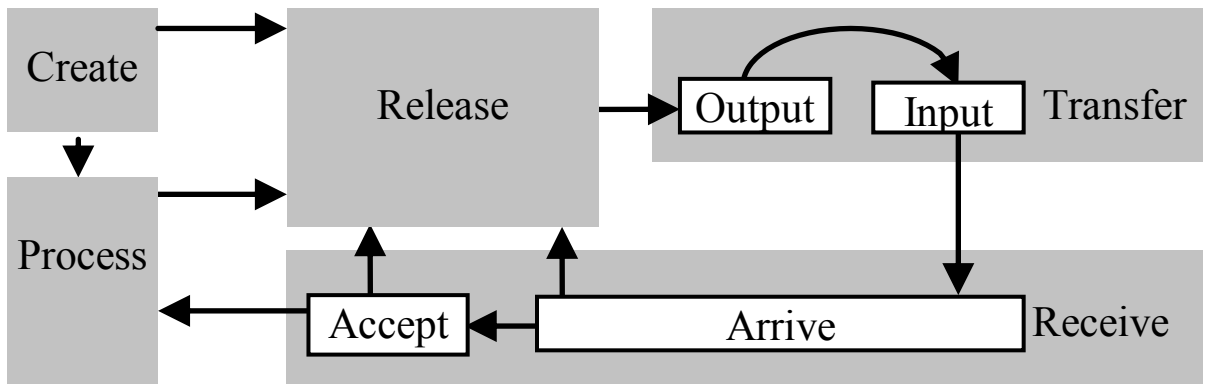

Fig. 2. Thinging machine.

- *Arrival*: A thing reaches a new machine.
- *Acceptance*: A thing is permitted to enter the machine. If arriving things are always accepted, then arrival and acceptance can be combined into the “receive” stage. For simplicity, this paper’s examples assume a receive stage exists.
- *Processing* (change): A thing undergoes a transformation that changes it without creating a new thing.
- *Release*: A thing is marked as ready to be transferred outside of the machine.
- *Transference*: A thing is transported somewhere outside of the machine.
- *Creation*: A new thing is born (created) within a machine. A machine creates in the sense that it finds or originates a thing; it brings a thing into the system and then becomes aware of it. Creation can designate “bringing into existence” in the system because what exists is what is found. Additionally, creation does not necessarily mean existence in the sense of being alive.

Creation in a TM also means appearance in the system. Appearance here is not limited to form or solidity but also to any sense of the system’s awareness of the new thing. Even nominals (which have no existence except in names) may be things that appear in the system model.

In addition, the TM model includes memory and triggering (represented as dashed arrows), or relations among the processes’ stages (machines), for example the process in Fig. 2 triggers the creation of a new thing.

The TM in Fig. 2 can be specified in a texual language (which we call TM language), wherein the arrows are represented by dots. For example, the following shows the different flows in Fig. 2 in this TM language:

*Flow.Create.release.transfer.output*
*Flow.Create.process.release.transfer.output*
*Flow.Transfer.input.receive.arrive.release.transfer.output*
*Flow.Transfer.input.receive.arrive.accept.release.transfer.output*
*Flow.Transfer.input.receive.arrive.accept.process.release.transfer.output*

The period is used to denote flow or containment. We use “-->” to indicate triggering.

## 3.2 Example

Fig. 3 shows the TM model of y = 10/x, x ≠ 0. The whole diagram is a thimac. The constraint x ≠ 0 is integrated into the model as a subthimac (the dark box). In Fig. 3, the value of x (as an independent variable) flows (circle 1) to the calculating subthimac, where it is received (2) and sent to the constraint subthimac (3). There, it is processed, and if it is not zero, (4) it flows back (5) to the calculating thimac that retrieves the constant 10 (6). Both values are processed (division; 7) to trigger (dashed arrow) creation of the value of y (8).

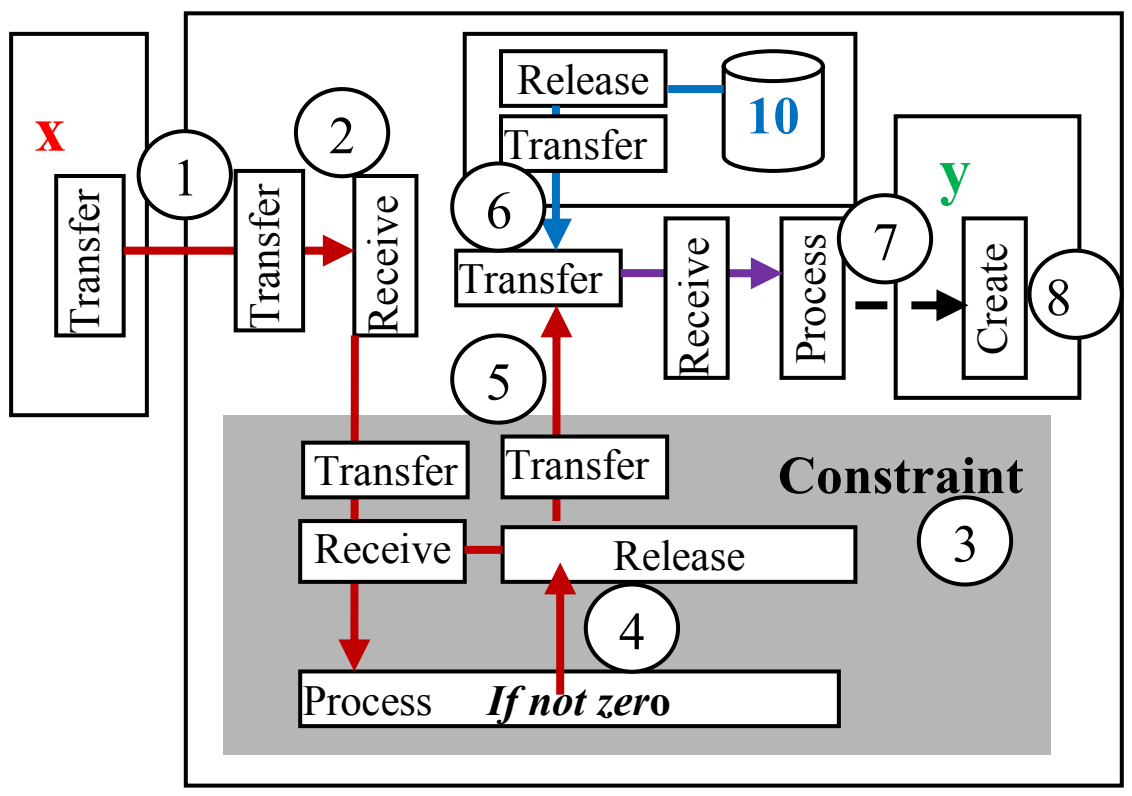

Fig. 3. The TM static model of y = 10/x, x ≠ 0.

To construct the dynamic model, we need to introduce the concept of an *event*. Consider the event *The value of x is inputted into the system*. Fig. 4 shows its representation in TM. It is simply a thimac that involves a time subthimac (the flow at the top). The *region* in the figure (dark box) is a subdiagram of the static model (Fig. 3). In general, an event may include other subthimacs (e.g., intensity or importance). For simplicity sake, we will represent each event by its region.

Thus, the dynamic model can now be represented in five events, as shown in Fig. 5.

Event A: The value of x is inputted.
Event B: The constant 10 is brought from its storage.
Event C: y is calculated.
Event D: The value of x is not zero.
Event E: The value of x is zero.

Fig. 6 shows the behavioral model of y = 10/x, x ≠ 0 in term of the event chronology. It embeds two types of behavior: A→D→B→C and A→E, where → indicates a sequence of events.

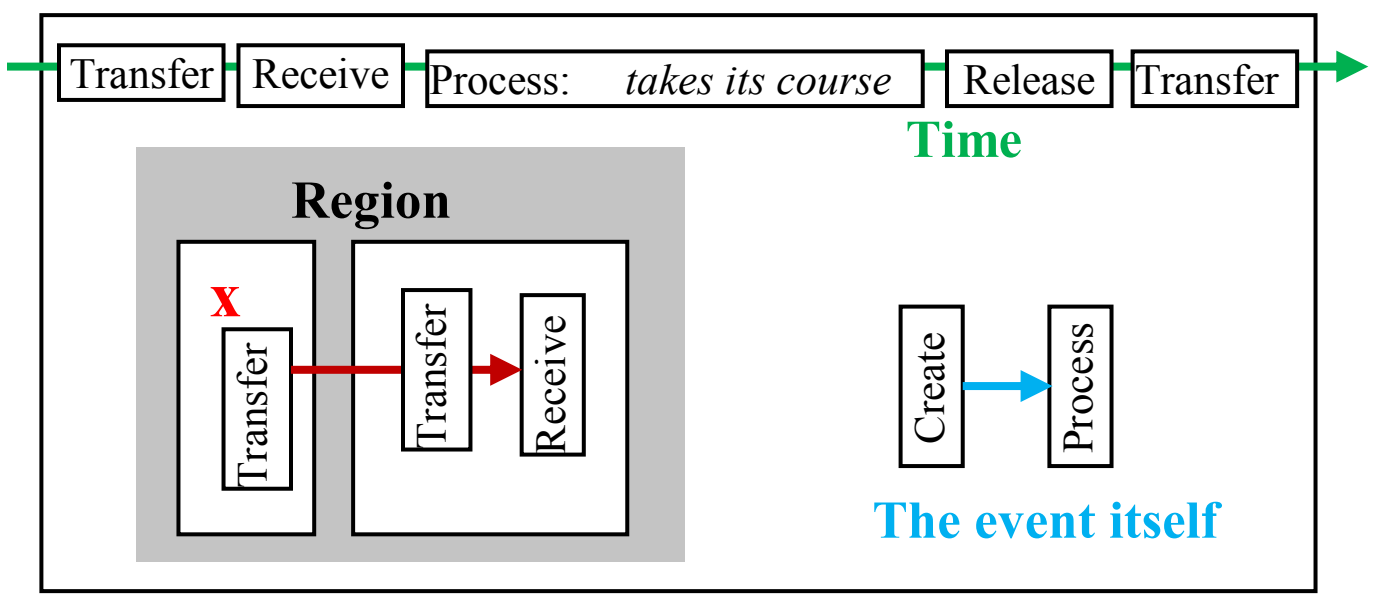

Fig. 4. Model of the event *x is inputted into the system.*

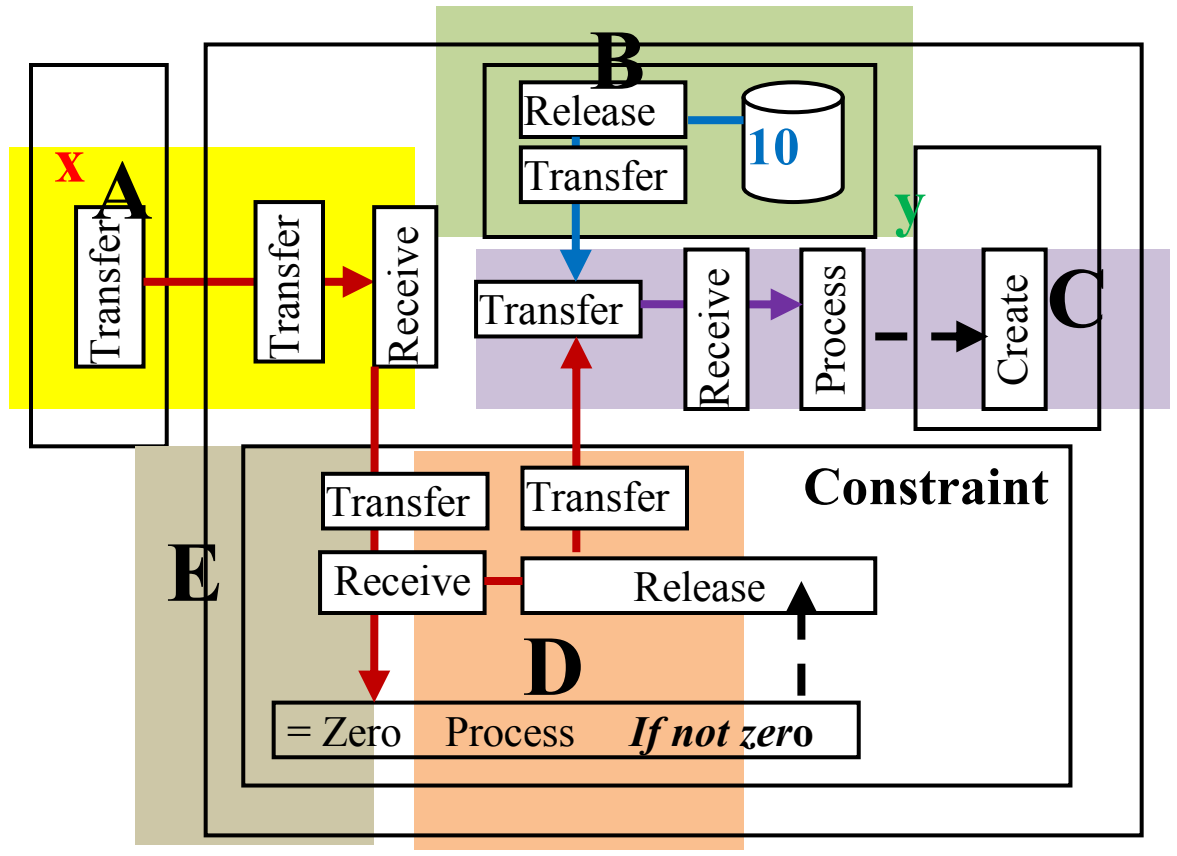

Fig. 5. The dynamic model of y = 10/x.

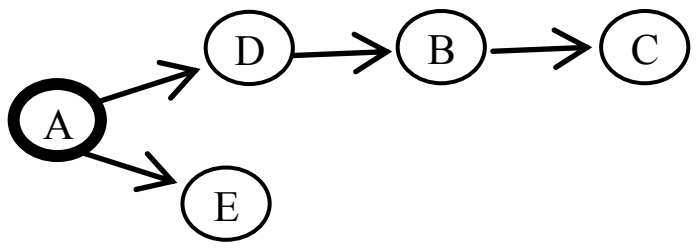

Fig. 6. The behavioral model of y = 10/x, x ≠ 0.

## 4. Case Study: Bridge–Island System

Abrial [3] modeled a physical/digital system that involves controlling cars on a narrow bridge that links the mainland to a small island. The system is equipped with two traffic lights (green and red) that control the entrance to the bridge at both ends. The system is equipped with sensors with two states—on or off—which are used to detect the presence of a car entering or leaving. The number of cars on the bridge and island is limited. The bridge is one-way or the other, not both at the same time.

Abrial [3] starts the model by developing a simple first version in which the various pieces of the traffic lights and sensors are introduced in subsequent refinements. The initial version includes a compound made of the bridge and the island together (see Fig. 7).

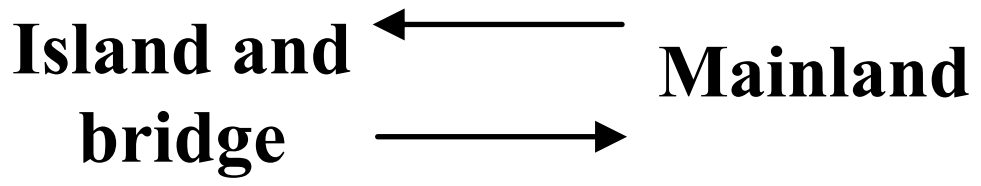

Fig. 7. Initial model of the mainland and the island/bridge (Adapted from [3]).

According to Abrial [3], "The idea is to take account initially of only a very few constraints. This is because we want to be able to reason about this system in a simple way, considering in turn each requirement." Abrial's [3] first task is to formalize the state of this simple version of the system and then formalize the two events of movement of cars between the mainland and the island–bridge.

In Event-B, the model's state is made up of two parts:

- The static part contains the definition and axioms associated with some constants. It contains *a single constant d*, which is a natural number denoting the maximum number of cars allowed on the island–bridge compound at the same time.
- The dynamic part contains the variables that are modified as the system evolves. It is made up of *a single variable n*, which denotes the actual number of cars in the island–bridge compound at a given moment.

The Event-B understanding of *static* is different from the TM use of this term. In TM, the number d is a thimac, while its value is subthimac in addition to its other subthimacs (e.g., its type).

The thimac d can be processed, released, transferred, and/or received. As shown in the example in the previous section, in TM, d is in the *static model*, which means that its operations may be specified but not as events in time. This is similar to the constant 10 in the example, which is a single value but is nevertheless released from its storage and transferred and then participates in calculating y. These are operations, not events, because they do not involve time.

In TM, describing d as a single-value thimac means that the value (and its type) is created (or initialized) once in the system. That is, no creation or transfer.receive occurs beside the original ones. The model in this case is static.

Fig. 8 shows the TM static model of this initial version of the example. It is important to emphasize the difference between the notions of an event in TM versus Event-B. To avoid redundancy, TM events of the bridge–island system are developed in the last refinement (sections 6-8). The flows between the mainland and the island–bridge (circles 1 and 2 in Fig. 8) are events in Events-B but are not events in TM.

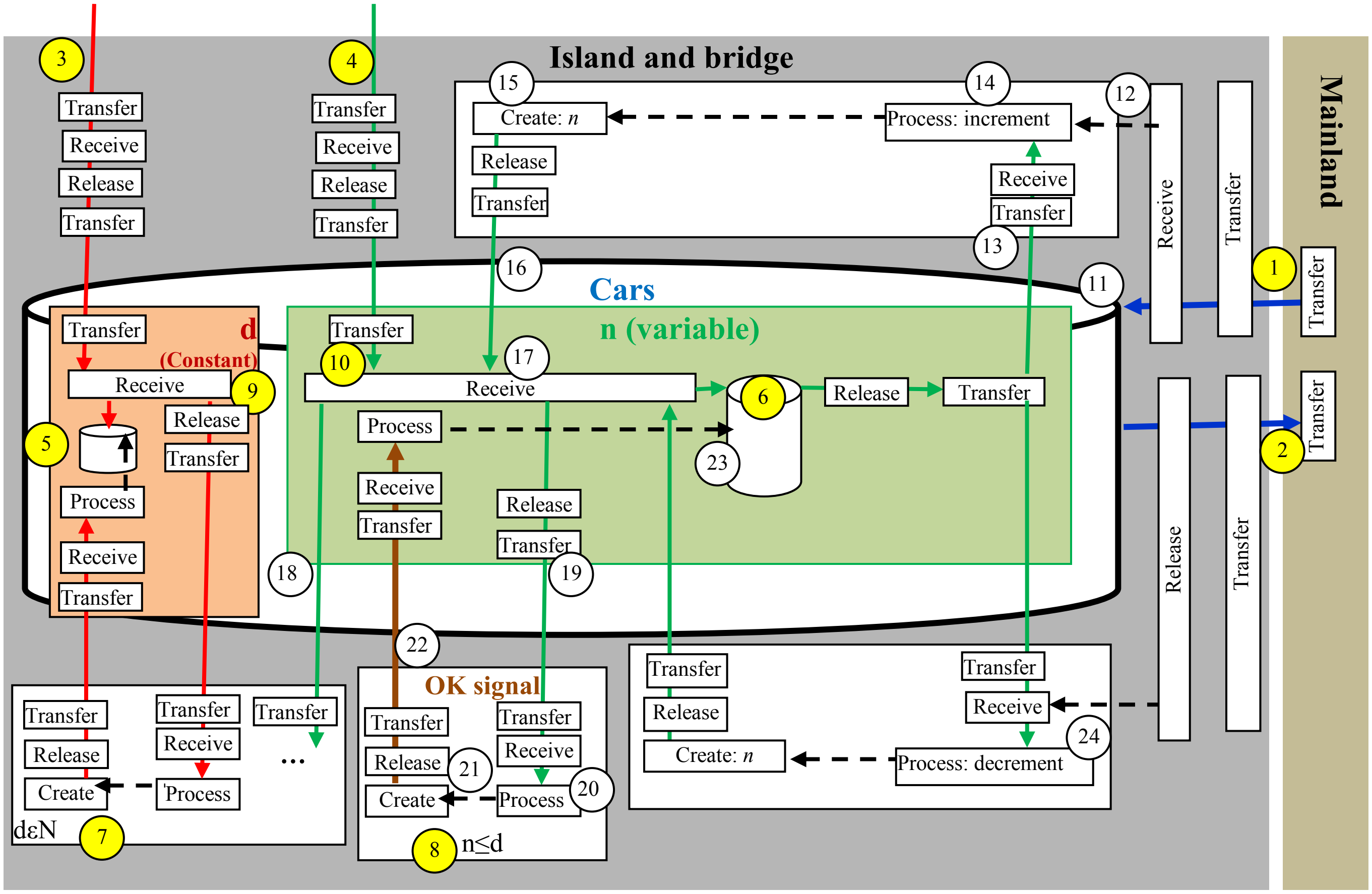

Fig. 8. The TM static model of the first version of the mainland and island–bridge example.

.
We assume d and n are initialized from outside (circles 3 and 4, respectively) at the start of the system and stored in the system (5 and 6 circles, respectively). The thimac dεN (7; we use dεN as a name) is responsible for the basic requirement of the axiom "d ε N": that d is a natural number. The thimac n≤d (8; we use n≤d as a name) is responsible for "n ≤ d." Because we assume d—the maximum number of cars on the island–bridge—is inputted (3) at the start of the system, then when this value is received (9), it is sent to dεN before it is accepted as a valid value and stored (5). Similarly, when n is inputted (4) and received, n is sent to dεN and n ≤ d before it is accepted as a valid value and stored (6). In TM, the thimacs dεN and n ≤ d are realizations of mathematical axioms. Any constraints on the system are treated as thimacs.

In the Event-B model, the static, dynamic, and behavioral models (in the TM sense) are mixed into a single mathematical representation. According to Abrial [3], a discrete transition component is made of a state and some transitions. Roughly, a state is defined by means of a number of variables, which might be any integers, pairs, sets, relations, functions, etc. In Event-B, the state might have any predicate expressed within the notation of first-order logic and set theory. Thus, "By putting all this together, a state can be simply abstracted to a set" [3].

An event in Event-B can be abstracted to a simple binary relation built on the state set. This relation represents the connection between two successive states, considered just before and just after the event's "execution". An event can be split into two parts: the guards and the actions. An action is an assignment to a state variable.

> A guard is a predicate, and all the guards conjoined together in an event form the domain of the corresponding relation. The actions of an event are supposed to be 'executed' simultaneously on different variables. Variables that are not assigned are unchanged. [3]

As we can see, the Event-B model has completely different conceptualizations of events, state, dynamism, behavior, etc., than TM does. Accordingly, we re-model the same system (the Event-B model in [3]) to contrast the two methodologies side-by-side.

In the TM static model of Fig. 8, the flow of cars between the mainland and the island–bridge (circles 1 and 2) causes (triggers) the updating of n (6). (Here, we mix the physical and the digital systems.) For example, the flow of a car to the island–bridge (11) triggers (12) the "pulling" of n (13) to be incremented by 1 (14). The new value of n (15) flows (16) to its thimac, where, upon arrival (17) and prior to storing it, it is sent to dεN (18) and n ≤ d (19) to check its new value. Note that, to simplify the diagram, we did not completely draw the flow from 18 since it is similar to the neighboring flow of d in dεN. The flow of n to n ≤ d is received and processed (20; that it satisfies the axiom n ≤ d) to send a signal (21) that the new value of n is acceptable (22); hence, this value is stored (23).

A similar explanation can be written when a car leaves the island–bridge (2) and n is updated by decrementing it (24).

## 5. First Refinement: One-Way Bridge

In Abrial's [3] first refinement of the initial model, the bridge is a one-way bridge and the constant d remains, but the variable n is now replaced by three variables: *a*, *b*, and *c*. Variable *a* denotes the number of cars on the bridge and going to the island, variable *b* denotes the number of cars on the island, and variable *c* denotes the number of cars on the bridge and going to the mainland, as illustrated in Fig. 9. The mathematical model continues to include the axioms:

$\mathbf{a} + b + c = \mathrm{d}$

$a = 0$ or $c = 0$

To save space and as we are going to present a fully detailed model of the system in the next section, we only show a partial view of the TM representation. Fig. 10 shows this TM static representation of this first refinement model. Note that due to space considerations, the diagram still does not separate the physical and digital thimacs.

There are three main physical thimacs: the mainland, bridge, and island. The bridge has two thimacs: one is the (physical) cars moving from the mainland to the island, and the other is the cars moving in the opposite direction.

- Assume that the bridge is initially open for cars coming from the mainland. A car moves from the mainland to the bridge (1), which triggers (2) retrieval of the (information system) value *a* and it being incremented (3). The new value is processed (4).

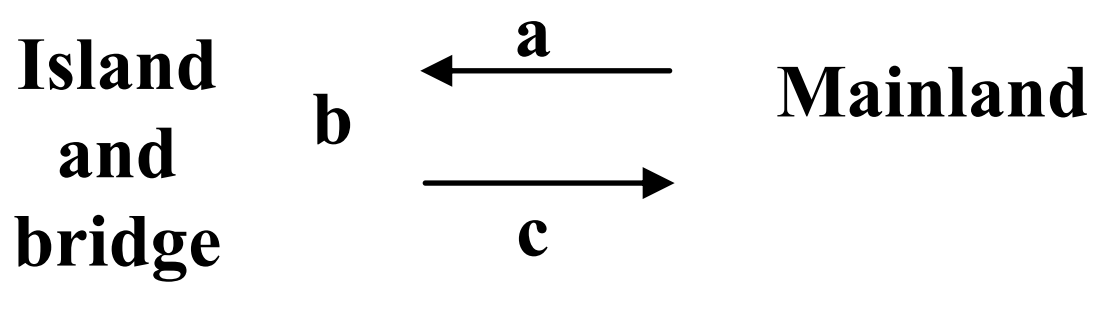

Fig. 9. Variables in the first version of the mainland and the island–bridge (Adapted from [3]).

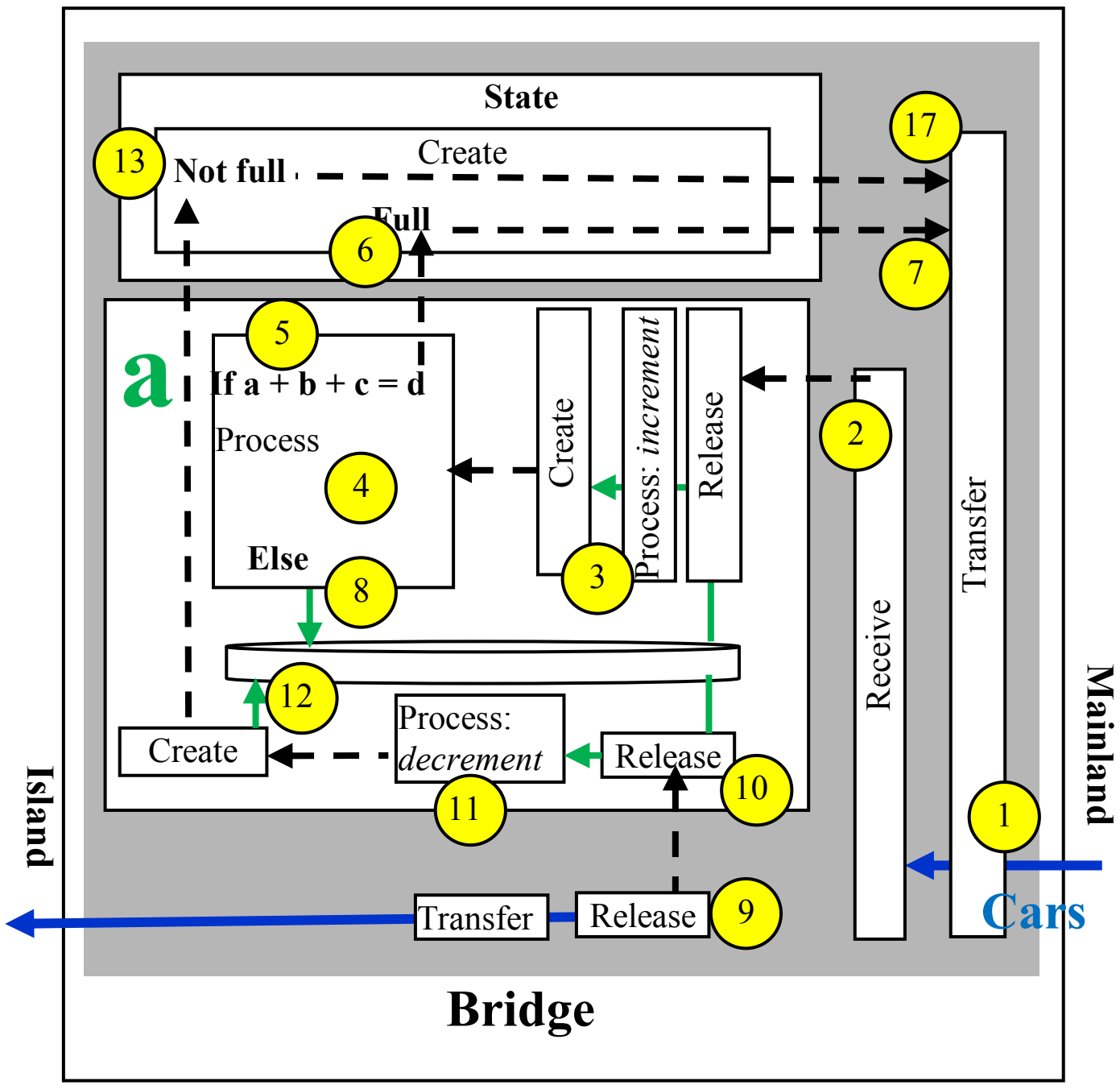

Fig. 10. Partial view of the TM static model of the first version of the mainland and island–bridge example.

- If $a + b + c =$ d (5), then the state (of the bridge) is full (6), and a close signal is sent to the gate (transfer) from the mainland (transfer; 7). For simplicity, this close signal is represented by a triggering. We assume the value of $a$ is initialized to zero at the start of the system.
- If $a + b + c <$ d (8; else, in the diagram), then the new value is stored.

- A car moves from the bridge to the island (9). This triggers (10) retrieving of the (information system) value of $a$, decrementing it (11), and
  - Storing the new value (12) in a.
  - Triggering of the gate's state (transfer) from the mainland to "not full" (13), which will open it if it was closed (14).

- Similar descriptions can be given for a car moving from the island to the bridge and for a car moving from the bridge to the mainland.
  Note that all of the thimacs in the initial model that check for values, nεN, are preserved in the first revision model, but for simplicity, they are not shown in Fig. 10.

## 6. Second Refinement: One-Way Bridge

In this last refinement, the sensors are introduced, which detect the physical presence of cars entering or leaving the bridge. These sensors are situated on each side of the road and at both extremities of the bridge. Additionally, traffic lights are installed at both ends of the bridge. Fig. 11 shows the static TM model of this refinement.

### 6.1 Bridge Traffic: From the Mainland/To the Island

**Traffic from the mainland**: A car enters the bridge (2). This triggers the sensor (3), which in turn triggers (4) to update ***a*** (5). The sensor also turns the traffic light to red, to prevent the entry of any additional cars while the current car is being processed. Note that the box labeled ***a*** is a module in the information system that maintains and stores the latest value of ***a***. This involves retrieving (releasing from storage) ***a*** (6) and incrementing it (7). The resulting value is examined (8), and accordingly,

- If $a$ (new value) $+ b + c =$ d, then ***a*** is stored, the bridge is full (cannot accept more cars; 9), and the traffic light is turned to red (10).
- Otherwise, the new value of ***a*** is stored (11), and the light is turned green to permit a new car to enter the bridge from the mainland.

**Traffic out to the island**: A car leaves the bridge (12), which triggers the sensor (13) to trigger ***a*** to be retrieved (14) and decremented (15). Additionally,

- A is stored (16).
- The state of the bridge is now "not full" (17); hence, the light is turned green (1).

### 6.2 Island Traffic: From Bridge/To Bridge

**Traffic in from the bridge**: On the island side, assuming the light (18) is on, a car enters the island from the bridge (19) to trigger the sensor (20), which triggers the retrieval of ***b*** (21). As explained before, the sensor also turns the light red. The value b is incremented (22) to be processed (23) as follows:

- If $a + b + c =$ d, then ***b*** is stored, the island is full, (24) and the light is turned red (25). Note that if the light is already red, then the color is not changed.
- Else (27), the new value of b is stored (28), and the light is turned on (29 and 18).

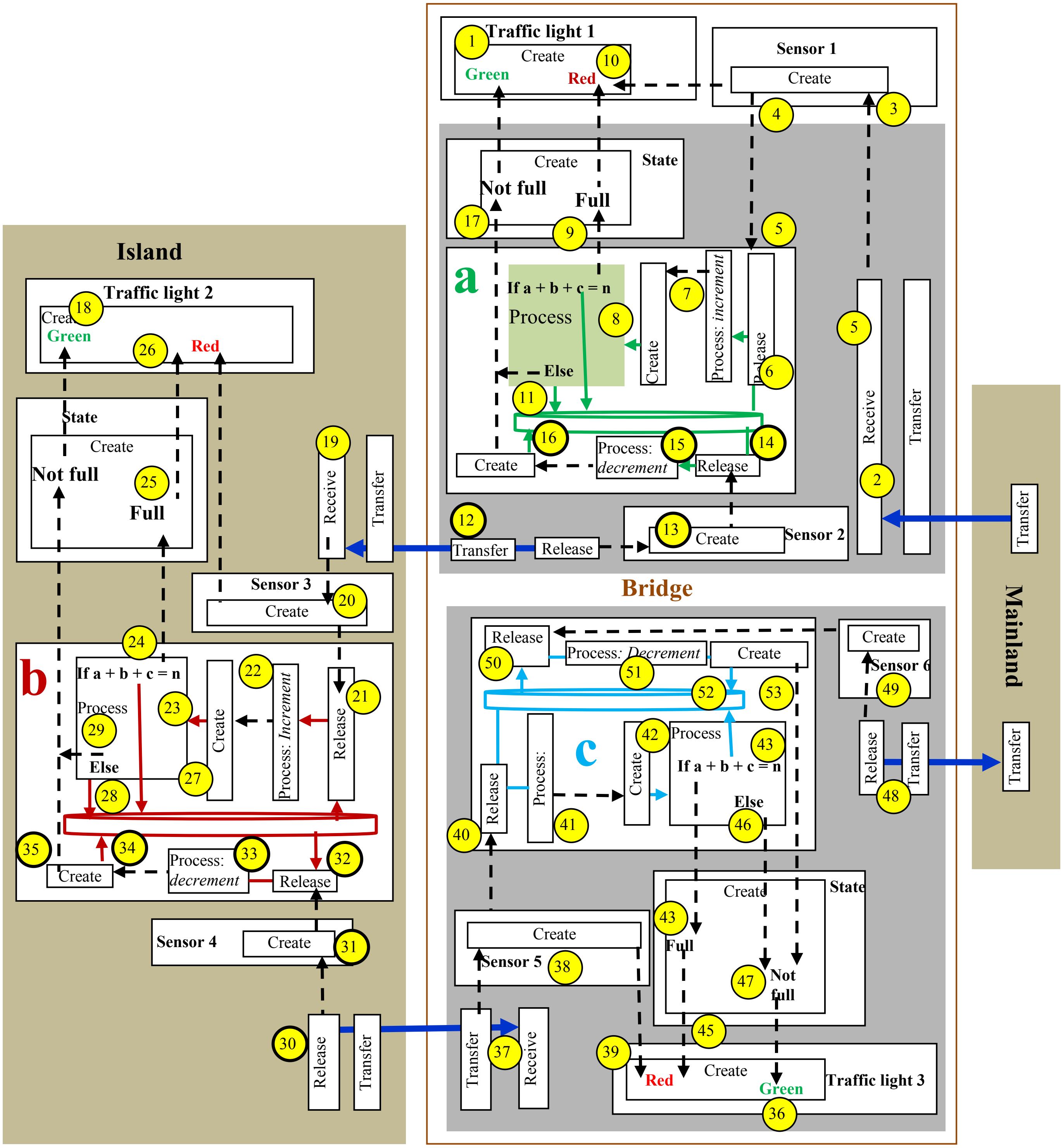

Fig. 11. The TM static model of the first version of the mainland and island–bridge example.

**Traffic to the bridge**: If a car leaves the island (assuming the light on the bridge side is green, as will be explained when describing the bridge traffic from the island to the mainland), then releasing the car (30) will trigger the sensor (31), which triggers the retrieval of *b* (32) and decremented (33). The new value is stored (34), and the light is turned on (35) to indicate that a space is available in the island.

### 6.3 Bridge Traffic: From Island/To Mainland

**Traffic out to the bridge**: Assuming that the traffic light on the bridge facing the island is green (36), a car enters the bridge (37) to trigger the sensor (38), which in turn turns the light red (39). Additionally, the sensor triggers the retrieval of ***c*** (49), which is processed (41) to be examined (42) as follows.

- If $a + b + c = d$, then ***c*** is stored, the bridge is full (43), and the light is turned red (45).
- Else (46), the bridge is full (47), and the light is turned red (36).

**Traffic to the mainland**: To simplify the figure, we ignore the traffic light on the mainland facing the bridge because this requires the mainland box to be expanded to the right of Fig. 11. Accordingly, we start with a car leaving the bridge to the mainland (48). This triggers the sensor (49), which triggers the retrieval of ***c*** (50). The value of ***c*** is decremented (51) and stored (52), and the light is turned to green (53).

## 7. Second Refinement: Dynamic Model

Fig. 12 shows the dynamic model that corresponds to the static representation in Fig. 11. Twenty-nine events are identified as follows.

**Events in the bridge (traffic from the mainland to the island)**

Event 1 ($E_1$): The traffic light is green.
Event 2 ($E_2$): A car enters the bridge.
Event 3 ($E_3$): The sensor is triggered.
Event 4 ($E_4$): The light is turned red.
Event 5 ($E_5$): ***a*** is retrieved, incremented, and processed.
Event 6 ($E_6$): $a + b + c$ is equal d.
Event 7 ($E_7$): ***a*** is stored
Event 8 ($E_8$): (Else: $a + b + c < d$)
Event 9 ($E_9$): A car leaves toward the island.
Event 9 ($E_{10}$): The sensor is triggered.
Event 10 ($E_{11}$): ***a*** is decremented, and a new ***a*** is created.

**Events in the island (traffic from the bridge and to the bridge)**

Event 12 ($E_{12}$): The traffic light facing the bridge is green.
Event 13 ($E_{13}$): A car enters the island.
Event 14 ($E_{14}$): The sensor is triggered.
Event 15 ($E_{15}$): The light is turned red.
Event 16 ($E_{16}$): ***b*** is retrieved, incremented, and processed.
Event 17 ($E_{17}$): a + b + c = d.
Event 18 ($E_{18}$): (Else: a + b + c < d).
Event 19 ($E_{19}$): ***b*** is stored
Event 20 ($E_{20}$): A car enters the bridge.
Event 21 ($E_{21}$): The sensor is triggered.
Event 22 ($E_{22}$): ***b*** is decremented, and a new ***b*** is created.

**Events in the bridge (traffic from the island and to the mainland)**

Event 23 ($E_{23}$): The traffic light facing the island is green.
Event 24 ($E_{24}$): A car enters the bridge.
Event 25 ($E_{25}$): The sensor is triggered.
Event 26 ($E_{26}$): The light is turned red.
Event 27 ($E_{27}$): ***c*** is retrieved, incremented, and processed.
Event 28 ($E_{28}$): $a + b + c = d$.
Event 29 ($E_{29}$): (Else) $a + b + c < d$, then ***c*** is stored.
Event 30 ($E_{30}$): ***c*** is stored
Event 31 ($E_{31}$): A car leaves toward the mainland.
Event 32 ($E_{32}$): The sensor is triggered.
Event 33 ($E_{32}$): ***c*** is decremented, and a new ***c*** is created.

These events provide us with a tool to construct different chronologies of events. We will specify each behavior according to the three areas: the bridge to the island, the island, and the bridge from the island.

## 8. Second Refinement: Behavioral Model

Fig. 13 shows the behavior of the bridge system that receives cars from the mainland and sends them to the island. It includes two streams of events: one starts when a car enters (events 1 and 2), and the other starts when a car leaves (event 9). As mentioned previously, this system involves the physical cars, traffic lights, and sensors in addition to the information system.

Similarly, Figs. 14 and 15 show the behavior of the system in the island and in the bridge leading to the mainland. Note the similarity of the three subsystems' behaviors. Fig. 16 shows the general behavior when all types of behaviors are connected together in terms of the chronology of events.

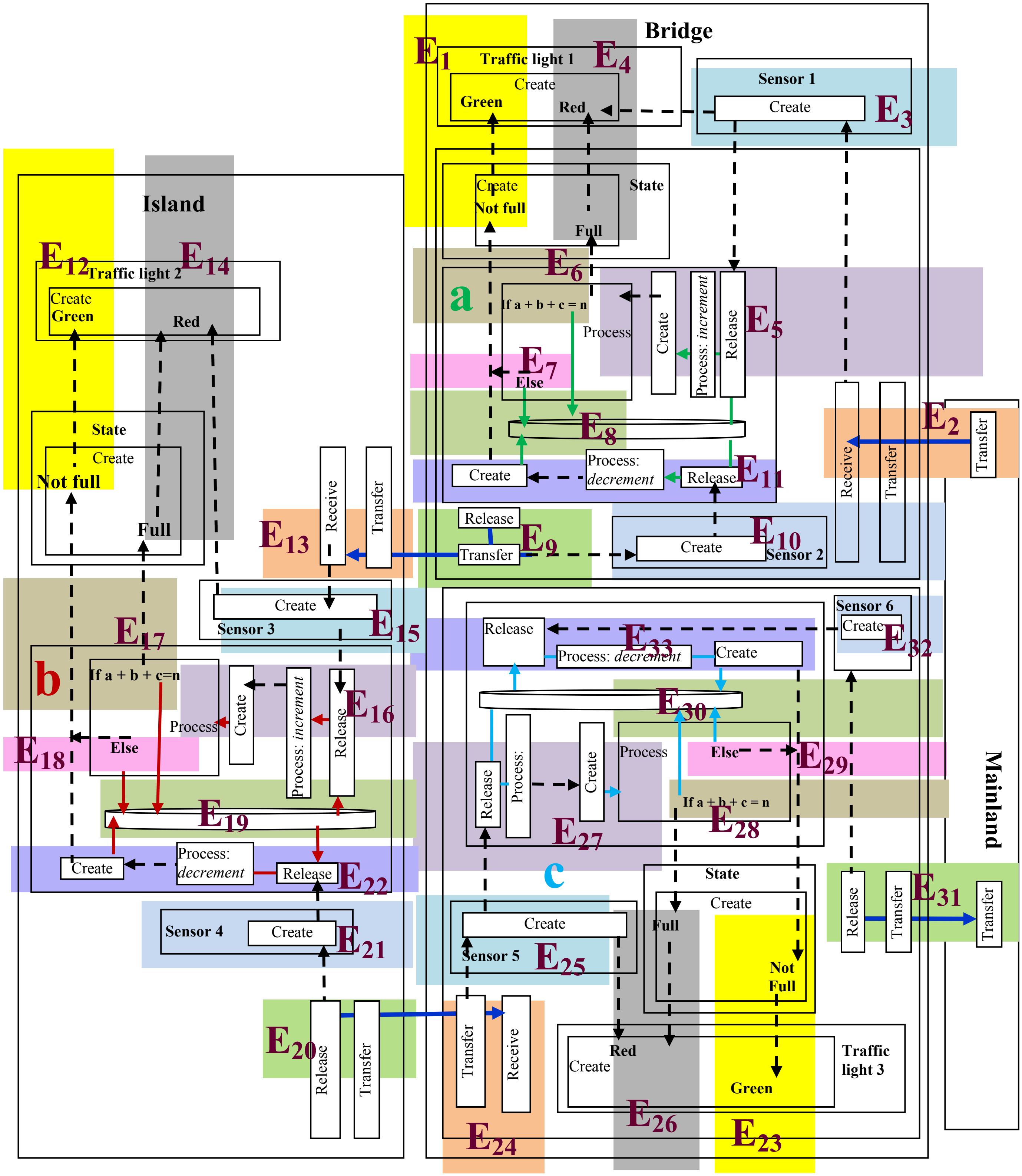

Fig. 12. The dynamic model of the bridge–island system.

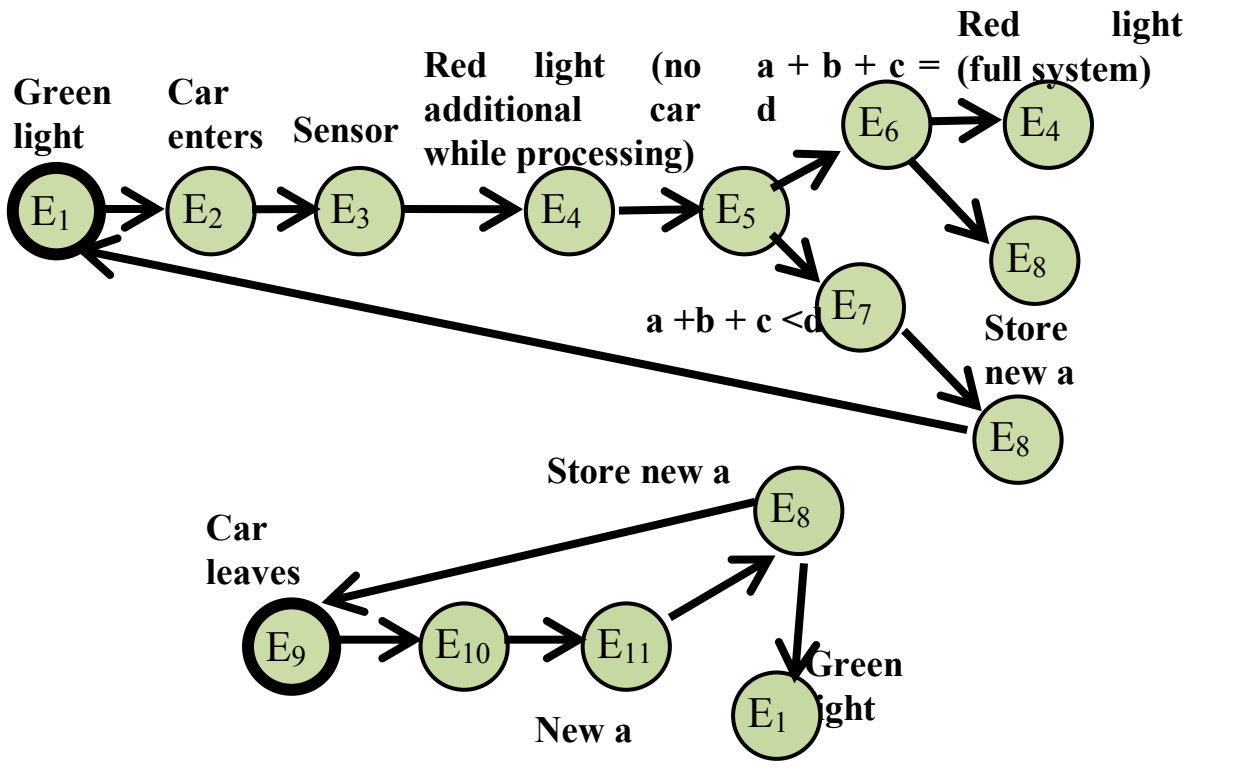

Fig. 13. Two behaviors in the bridge where cars are coming from the mainland (upper) and/or leaving toward the island (lower).

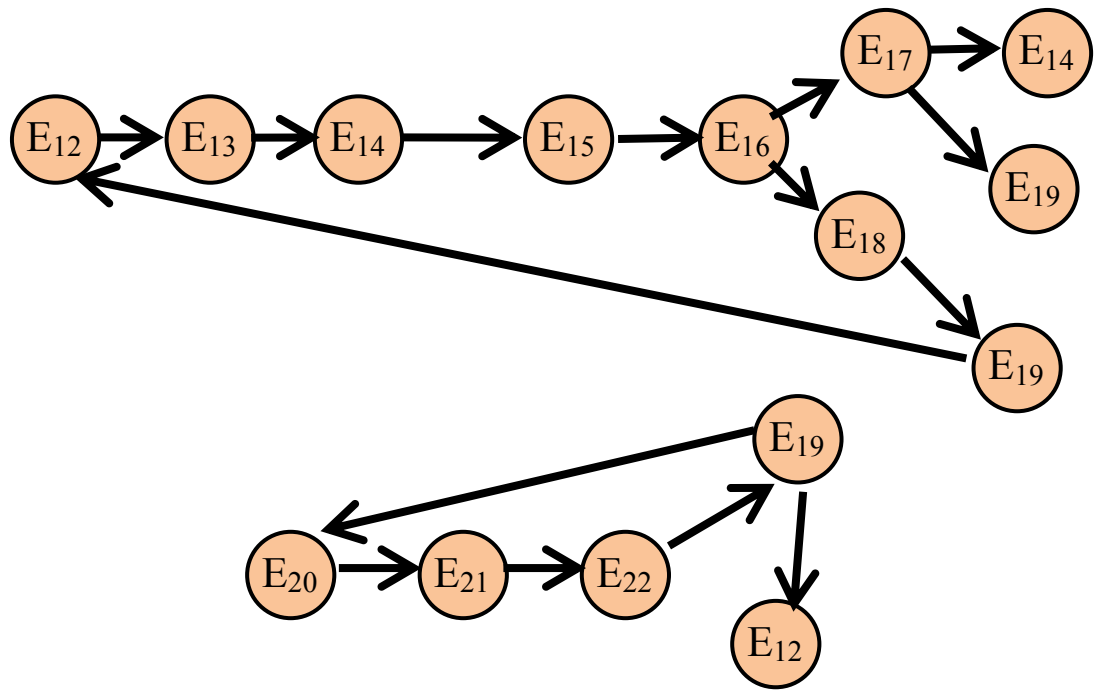

Fig. 14. Two behaviors in the island where cars are coming from the bridge (upper) and/or leaving toward the bridge (lower).

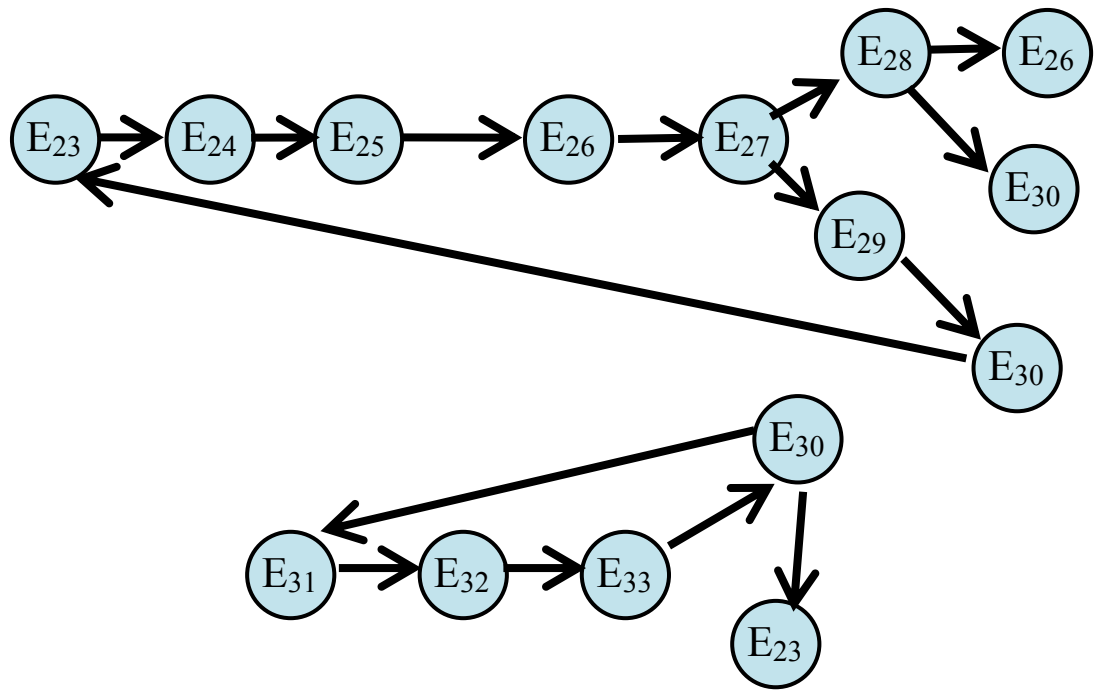

Fig. 15. Two behaviors in the bridge where cars are coming from the island (upper) and/or leaving toward the mainland (lower).

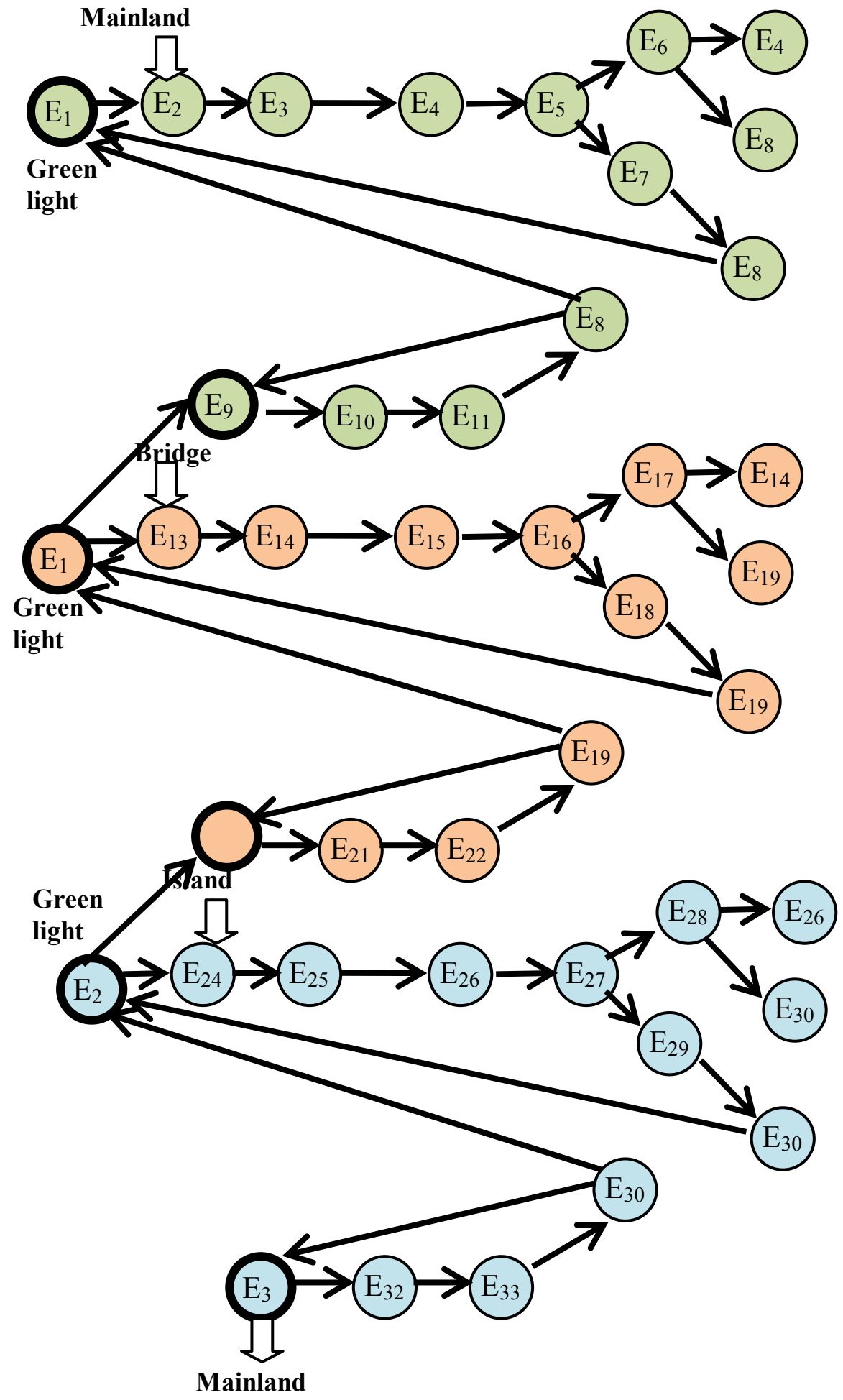

Fig. 16. The general behavior of the bridge–island system.

## 9. Digital System

Note that the chronology of events in the behavioral model is mandated by flows among the generic operations (e.g., a variable is released from storage and processed), which generates a release event that precedes a process event. However, if there is no flow (e.g., receiving a new car cannot generate flow in the sensor), then triggering is used to “force” a sequence among different streams of (disconnected) flows (e.g., receiving a car *triggers* a signal being created in the sensor). Thus, triggering is a mechanism in the static model for early enforcement of a certain chronology of events in the behavior model when there is no flow. The triggering can be implemented as a communication signal in further refinements of the behavior model to develop the control through an information system.

Fig. 17 is an example of such a development, in which the information is separated from the physical system. In addition, triggering in the static model (e.g., the sensor triggers updating *a*) is implemented as a communications signal (dotted boxes). The same information system can be used to control all three regions of the system, as shown in Fig. 18. Depending on which sensor is sending a signal, *a*, *b*, or *c* is used, assuming that d is a global variable.

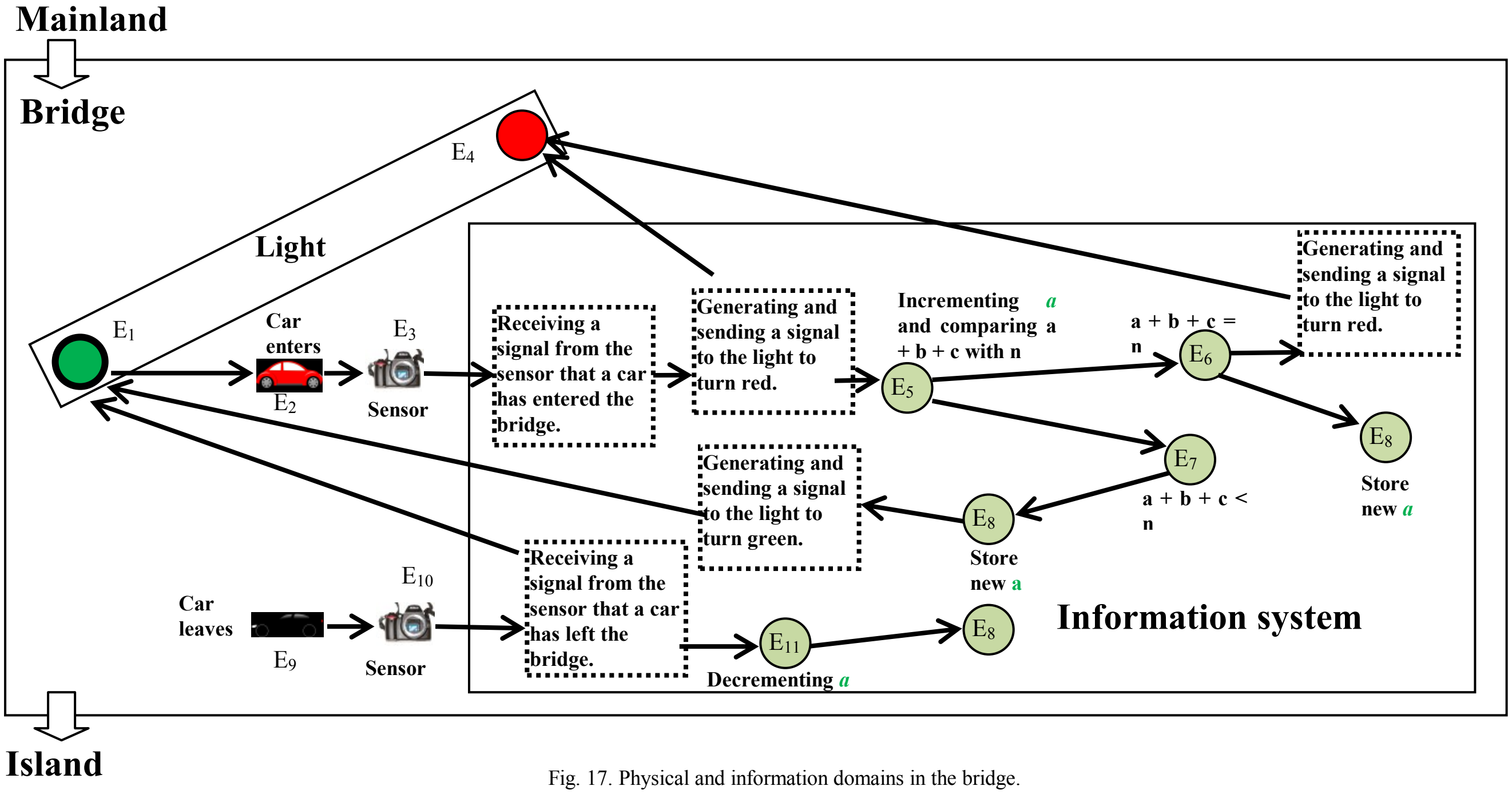

Fig. 17. Physical and information domains in the bridge.

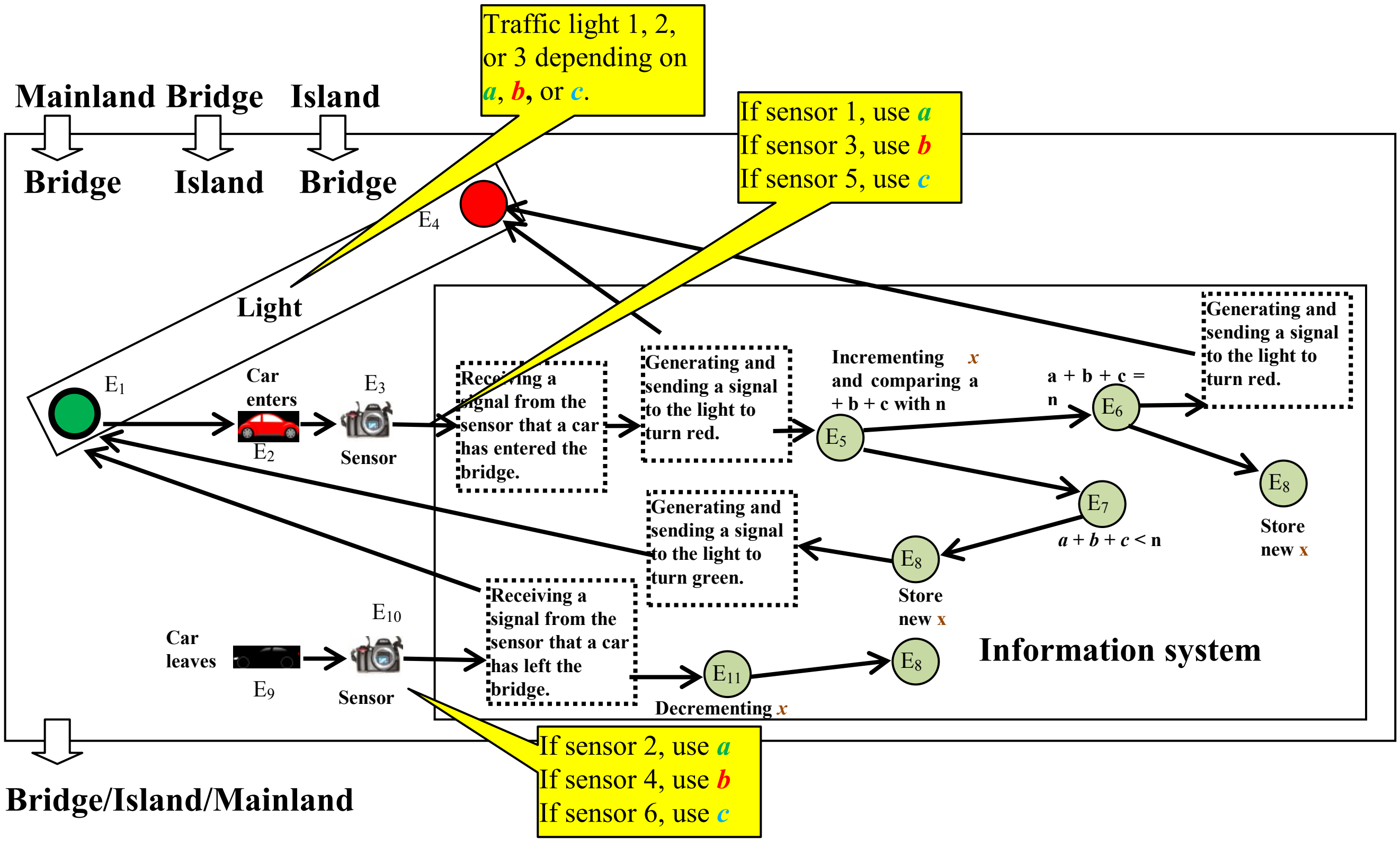

Fig. 18. The information system is activated according to the sensors.

## 10. Conclusion

In this paper, we examined two modeling methodologies—the formal Event-B and the diagrammatic TM—through a single case study of a bridge–island system. Contrasting the two diverse models for the same large problem is an inspiring venture. It has enhanced the advantages and limitations of the modeling experience. On one hand, Event-B is an attractively molded formalism of mathematically scattered notations with no focusing center that ties together the whole in a main structure. TM seems to be a holistic assembly with a recognizable center, but it has a voluminous form that needs much fastening elements (e.g., triggering). Event-B facilitates proving but seems impractical, at least for large systems. TM is easy to apply but seems difficult to maintain, at least for large systems. This contrasting process may be extended to many aspects in the two models. It seems that mixing formal and diagrammatic styles is a promising approach. This approach has already been adopted in Event-B through developing UML-B [7].